\documentclass[11pt,a4paper]{article}
\pdfoutput=1
\usepackage{jcappub}

\usepackage{color}

\usepackage{booktabs}

{\count255=\time\divide\count255 by 60 \xdef\hourmin{\number\count255}
  \multiply\count255 by-60\advance\count255 by\time
  \xdef\hourmin{\hourmin:\ifnum\count255<10 0\fi\the\count255}}

\newcommand{\impc}{\mathrm{Mpc}^{-1}}

\newcommand{\Planck}{{\emph{Planck}}}

\begin{document}

\title{The Knotted Sky II:  Does BICEP2 require a nontrivial primordial power spectrum?}

\author[a]{Kevork N. Abazajian,}
\author[b]{Grigor Aslanyan,}
\author[b]{Richard Easther,}
\author[b]{and Layne C. Price}

\affiliation[a]{Department of Physics, University of California at Irvine,
  Irvine, CA 92697\vspace{4pt} }
\affiliation[b]{Department of Physics, University of Auckland, Private Bag 92019, Auckland, New Zealand}

\emailAdd{kevork@uci.edu}
\emailAdd{g.aslanyan@auckland.ac.nz}
\emailAdd{r.easther@auckland.ac.nz}
\emailAdd{lpri691@aucklanduni.ac.nz}

\date{\today}

\abstract{
  An inflationary gravitational wave background consistent with BICEP2  is difficult to reconcile with a simple power-law spectrum of primordial scalar perturbations. Tensor modes contribute to the temperature anisotropies at multipoles with $l\lesssim 100$, and this effect --- together with a prior on the form of the scalar perturbations --- was the source of previous bounds on the tensor-to-scalar ratio. We compute Bayesian evidence for combined fits to  BICEP2 and \Planck\  for three nontrivial primordial spectra: a) a running spectral index, b)  a cutoff at fixed wavenumber, and c) a spectrum described by a linear spline with a single internal knot. We find no evidence for a cutoff, weak evidence for a running index, and significant evidence for a ``broken'' spectrum. Taken at face-value, the BICEP2 results  require two new inflationary parameters in order to describe both the broken scale invariance in the perturbation spectrum and the observed tensor-to-scalar ratio. Alternatively, this tension may be resolved by additional data and more detailed analyses. }
\maketitle

\section{Introduction}

The BICEP2 experiment \cite{Collaboration:2014wq,Collaboration:2014xx} has reported a detection of primordial B-modes in the cosmic microwave background (CMB).\footnote{B-mode polarization from CMB lensing was detected earlier by the POLARBEAR experiment \cite{Ade:2014afa}.} The most natural explanation for a B-mode signal is a stochastic background of long-wavelength gravitational waves, or tensor perturbations \cite{Seljak:1996gy}. This constitutes strong {\em prima facie\/} evidence for an inflationary phase in the early universe, the most widely-studied source for a stochastic background of gravitational waves.  If the observational data and theoretical explanation are confirmed,  the B-mode signal will provide unprecedented insight into the mechanism responsible for inflation.

The measured tensor-to-scalar ratio has a $68\%$ confidence-interval (CI) of $r=0.20^{+0.07}_{-0.05}$ and differs from zero with  a statistical significance of $5.9 \sigma$. However, the temperature data from \emph{Planck} \cite{Ade:2013xsa,Ade:2013zuv,Ade:2013uln}, SPT \cite{Hou:2012vk}, and ACT \cite{Das:2013wg}, combined with WMAP \cite{Hinshaw:2012aka} polarization, yields $r \lesssim 0.11$ at the $95\%$ CI,   in significant tension with the BICEP2 result. 

There are several potential explanations for this discrepancy. The first is that the BICEP2 analysis over-estimates the amplitude of the B-mode itself {{\cite{Mortonson:2014bja}}. The second possibility is that the primordial B-mode is accurately measured, but sourced by a mechanism unrelated to the standard assumptions for the primordial inflationary phase \cite{Seljak:1997ii,Pogosian:2007gi,JonesSmith:2007ne,Kobayashi:2010pz,Cook:2011hg,Senatore:2011sp,Barnaby:2012xt,Carney:2012pk,Contaldi:2014zua,Czerny:2014wua}. Conversely, existing CMB data may have been misanalysed, although this appears unlikely given the  agreement of \emph{Planck} with WMAP at large and intermediate scales and with ACT and SPT at small scales.

Another suggestion is that the large-scale scalar power spectrum is suppressed relative to that predicted by the best-fit $\Lambda$CDM scenario. Pre-BICEP2 constraints on the inflationary gravitational wave background were driven primarily by the contribution of tensor modes to the temperature-temperature (TT) anisotropies. This is illustrated in Fig.~\ref{cl_tt_fig}, which  shows the contribution to $C_l^{TT}$ at low-$l$ from a tensor background with $r=0.2$. The tensor contribution to individual $C_l^{TT}$ is small, but systematically increases the TT multipoles for all $l\lesssim 100$.   However, to constrain the primordial tensor background using this signal we must have an independent estimate of the  contributions $C_l^{TT}$  from scalar perturbations alone.  Moreover, the  measured low-multipole $C_l^{TT}$ typically lie below the best fit values for simple power-law spectra, so the inclusion of a tensor background is likely to reduce the likelihood relative to $r=0$.  Fig.~\ref{cl_tt_fig} also shows a sample angular power spectrum derived from a primordial scalar spectrum with a sharp cut-off in power at a comoving wavenumber $k=0.002 \, \impc$.  This particular scenario has a very low likelihood relative to the \emph{Planck} and WMAP datasets, but provides an extreme illustration of how a scalar spectrum with a cutoff  could compensate for a tensor contribution to the $C_l^{TT}$ for $l\lesssim 100$.

In this paper we focus on the implications of the BICEP2 result for the scalar power spectrum $\Delta^2(k)$,  performing joint analyses of the BICEP2 and \emph{Planck} datasets.   We consider three possibilities: (i) a running spectral index, (ii) a sharp cut-off in power at scale $k_\mathrm{cut}$, and (iii) a discontinuity in the spectral index at scale $k_\mathrm{knot}$. The latter two scenarios are implemented via the algorithm described by us in Ref.~\cite{Aslanyan:2014mqa}.    

The BICEP2 analysis  \cite{Collaboration:2014wq,Collaboration:2014xx} presents joint constraints from BICEP2 and \Planck\ with a running index, but focusses primarily on the polarization and B-mode amplitude and does not discuss the issue in detail.  We reproduce the BICEP2 constraints on a running index, and compute Bayesian evidence (relative to $\Lambda$CDM) of $\Delta \log Z=1.1$ for the running case.  The cutoff spectrum does not give a significant improvement, since it suppresses the scalar power by a factor far larger than the corresponding increase in power due to tensor contributions.  Finally, a break in the spectral index --- implemented as a splined $P(k)$ with a single ``interior'' point at an arbitrary amplitude and location --- provides the best fit to the data. With a broad, uninformative prior we find the change in the logarithmic evidence ratio is $\Delta \log Z=1.6$.   However, using a prior that includes information from our investigation of the \Planck\ data alone, which disfavors knots at $k_\mathrm{knot} \gtrsim 10^{-2} \, \impc$ or dramatic changes in amplitude, we infer an evidence ratio of $\Delta \log Z=3.1$, a ``significant'' to ``strong'' detection according to conventional model selection criteria.

Consequently, this analysis would suggest that BICEP2 has actually made {\em three\/} significant discoveries about  inflation: the first is to confirm its existence, and the second is to show that it took place at a relatively high energy scale. Thirdly, these results also suggest that  the primordial scalar spectrum has a nontrivial structure which is \emph{inconsistent} with  simple models of inflation.   Alternatively, the tensor spectrum may differ from the ``standard'' inflationary form, due to either a variant model of inflation \cite{Miranda:2014wga} or a mechanism that is independent of inflation  \cite{Seljak:1997ii,Pogosian:2007gi,JonesSmith:2007ne,Senatore:2011sp,Contaldi:2014zua}. Of course, the conservative explanation of our findings is that they point to tension between the \Planck\ and BICEP2 datasets which will be resolved by more complete analyses and/or additional data.  

\begin{figure}
\centering
\includegraphics[width=1.0\textwidth]{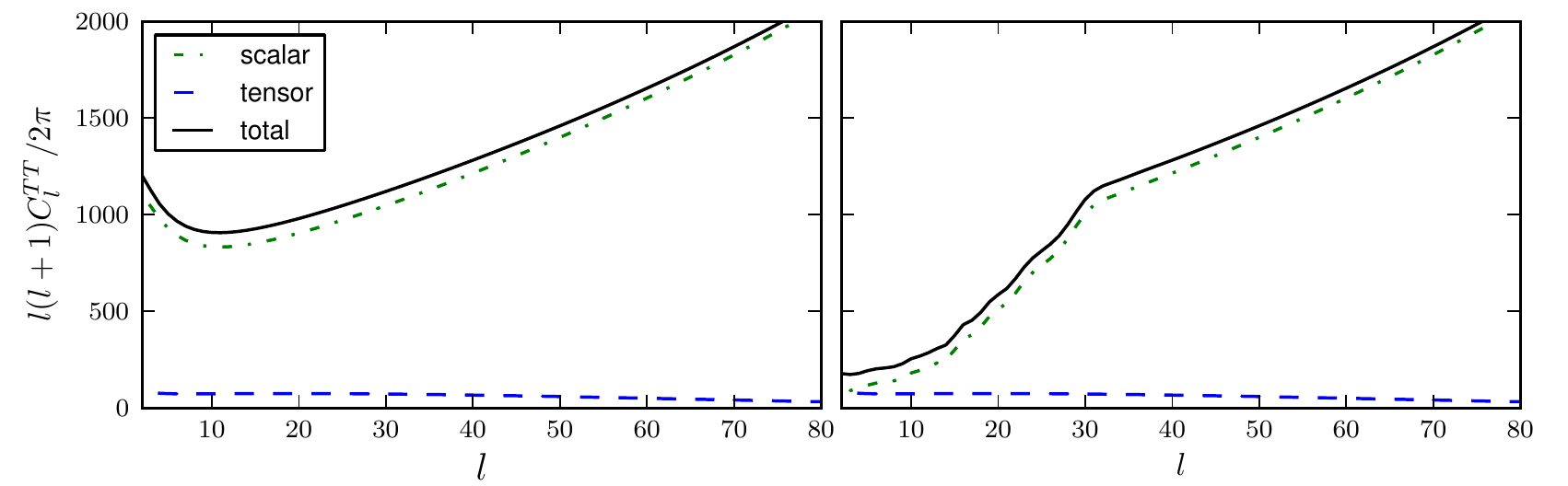}
\caption{\label{cl_tt_fig} The contribution to the (lensed) TT power spectrum from scalar and tensor modes ($r=0.2$). The left panel shows a standard power law primordial power spectrum with typical values of the cosmological parameters. The right panel shows the same primordial power spectrum but cut-off below $k=0.002 \, \impc$ (corresponding to $l\sim30$).
}
\end{figure}

\section{Method}

\subsection{Likelihoods, priors, and Bayesian evidence}

We combine the $B$-mode results from BICEP2 \cite{Collaboration:2014wq,Collaboration:2014xx} with  temperature and lensing data from \emph{Planck} \cite{Ade:2013xsa}. We use the \emph{Planck} likelihood code \cite{Collaboration:2013vc} with \textsc{Commander}, \textsc{CamSpec}, and lensing likelihood files for the data-likelihood evaluation.  We use the \textsc{COSMO++} library \cite{Aslanyan:2013ts} to combine the modified form of the primordial scalar power spectrum (Fig.~\ref{model_fig}) with the \emph{Planck} likelihood code and calculate the CMB angular power spectra with CLASS \cite{Lesgourgues:2011ud,Blas:2011vf}.  We employ multimodal nested sampling for parameter estimation and the computation of  evidence, using the publicly available code \textsc{MultiNest} \cite{Feroz:2007fi,Feroz:2008eb,Feroz:2013uq}.

We use \emph{model posterior probabilities} and \emph{Bayesian evidences} to compare the statistical significance of two competing reconstruction models.  This approach penalizes models with parameters for which the likelihood is large only in small regions of parameter space and protects against overfitting the scalar spectrum with too many knots or bins.  This approach has  previously been employed for power spectrum reconstruction \cite{Bridges:2005br,Bridges:2006zm,Bridges:2008ta,2012JCAP...06..006V,Vazquez:2011xa,Vazquez:2013dva} and gives conservative and robust assessments of possible physical features in the data.

We use the posterior probability $P(\mathcal M \, | \, \mathrm{Data})$ to assess the statistical significance of a model $\mathcal M$.  If   two models $\mathcal M_i$ and $\mathcal M_j$ have the same prior probability,   Bayes' theorem yields the relative betting odds between the models via the Bayes factor
\begin{equation}
  B_{ij} = \frac{P(   \mathrm{Data}\, | \, \mathcal M_i)}{P(   \mathrm{Data}\, | \, \mathcal M_j)} \, ,
  \label{eqn:bayes_factor}
\end{equation}
where the Bayesian evidence (\emph{marginalized likelihood}) is
\begin{equation}
  P(   \mathrm{Data} \, | \,\mathcal M_i) = \int P(\theta \, | \, \mathcal M_i ) \, \mathcal L(\mathrm{Data} \, | \, \theta) d\theta
  \label{eqn:evidence}
\end{equation}
for the model parameters $\theta$.  Here, $\mathcal L(\mathrm{Data} \, | \, \theta)$ is the data-likelihood and $P(\theta \, | \, \mathcal M_i )$ is the parameter prior probability.  Equation~\eqref{eqn:evidence} is a function of our choice of prior, so to be conservative we allow the parameters that define the power spectrum features to vary over a wide range of values in order to thoroughly search the parameter space \cite{Easther:2011hj}. Because there is little likelihood for features at $k \gtrsim 10^{-3} \, \impc$, a model with more parameters must produce a very large improvement in the likelihood relative to a featureless scenario for evidence to yield ``betting odds'' that strongly support the more complex model. Consequently, we also examine the improvement in fit ($\Delta {\cal L}$) at the maximum likelihood point for each scenario.
 
\begin{table}
\renewcommand{\arraystretch}{1.5}
\setlength{\arraycolsep}{5pt}
\centering
\begin{tabular}{c | c c}
  \hline
  \hline
  $\log $ (\emph{Posterior Odds}) &  Jeffreys Scale & Cosmology Scale \\
\hline
0.0 to 1.0 &  \multicolumn{2}{c}{Not worth more than a bare mention}  \\
1.0 to 2.5 &  Substantial & Weak \\
2.5 to 5.0 &  Strong & Significant \\
$> 5$      &  Decisive & Strong \\
  \hline
  \hline
\end{tabular}
\caption{Rough guideline for Bayesian evidence interpretation with the Jeffreys scale \cite{Jeffreys:1961} and the more conservative ``cosmology scale'' from Ref. \cite{hobson2010bayesian}.  The posterior odds equal the Bayes factor when the prior models odds are equal.}
\label{jeffreys_table}
\end{table}

We use uniform priors for all of the standard cosmological parameters $\Omega_bh^2$, $\Omega_ch^2$, $h$, and $\tau$, as well as the $14$ ``nuisance'' parameters  in the \textsc{CamSpec} likelihood. For convenience, we set the prior probability distribution for $r$ to the posterior distribution obtained by BICEP2 \cite{Collaboration:2014wq,Collaboration:2014xx}, which is equivalent to the direct evaluation of the underlying likelihood with a uniform prior on $r$. {Consistent with  BICEP2, we assume a flat tensor power spectrum and a pivot scale of $k_*=0.05\,\mathrm{Mpc}^{-1}$ \cite{Collaboration:2014wq,Audren:2014cea}.} We   check that we recover the joint constraints on $n_s$ and $r$  and the marginalised posterior for the running index reported by BICEP.

\begin{table}
\renewcommand{\arraystretch}{1.5}
\setlength{\arraycolsep}{5pt}
\centering
\begin{tabular}{c | c | c }
  \hline
  \hline
  \multicolumn{3}{c}{Broad Priors} \\
  \hline
  Linear Spline & Cutoff & Running \\
  \hline
  $-6 < \log_{10} k_\mathrm{knot} < 0$ & $-6 < \log_{10} k_\mathrm{cut} < 0$ & $-0.1 < \alpha_s < 0.1 $ \\
  $-2 < \log (10^{10} \Delta_\mathrm{knot}^2) < 4$ & & \\
  \hline
  \hline
\end{tabular}

\vspace{1cm}

\begin{tabular}{c | c }
  \hline
  \hline
  \multicolumn{2}{c}{Informative  Priors} \\
  \hline
  Linear Spline & Cutoff \\
  \hline
  $-6 < \log_{10} k_\mathrm{knot} < -2$ & $ -6 < \log_{10} k_\mathrm{cut} < -2$ \\
  $2 < \log (10^{10} \Delta_\mathrm{knot}^2) < 4$ & \\
  \hline
  \hline
\end{tabular}
\caption{Priors for the non-standard scalar power spectrum: the location $k_\mathrm{knot}$ of the knot, the amplitude of the knot's dimensionless scalar power spectrum $\Delta^2$, the running of the spectral index $\alpha_s$, and the cutoff scale $k_\mathrm{cut}$.  The values for $r$ are drawn from the BICEP2 likelihood $\mathcal L_\mathrm{BICEP2}$.  All cosmological parameters not listed above, as well as the amplitudes at the endpoints, have the same priors as used in our previous analysis, Ref.~\cite{Aslanyan:2014mqa}.}
\label{priors_table}
\end{table}

\subsection{Non--power-law scalar spectrum}

We perform a complete marginalization of the \emph{Planck} temperature and lensing likelihood, varying the remaining $\Lambda$CDM variables and the full set of \emph{Planck} ``nuisance parameters.'' We calculate the  Bayesian evidence ratios and combine this with the parameter posterior probabilities to give a conservative model-selection guideline.  We use a scale-invariant primordial tensor spectrum and  draw $r$ from a prior defined by the BICEP2 posterior, as noted previously. 

We summarize all power spectrum priors in Table~\ref{priors_table}.  We reconstruct the primordial scalar power spectrum in the range $10^{-6} \, \impc < k < 1.0 \, \impc$, using a generalization of the ``knot-spline'' procedure, developed in Refs~\cite{Bridges:2005br,Bridges:2006zm,Verde:2008er,Bridges:2008ta,Peiris:2009ke,Bird:2010mp,2012JCAP...06..006V,Vazquez:2011xa,dePutter:2014vd,Aslanyan:2014mqa}.  This process is illustrated  in Fig.~\ref{model_fig}. A complete discussion is given in  Ref.~\cite{Aslanyan:2014mqa}, but it can be summarised as follows: 

\begin{enumerate}
  \item  Fix the endpoints at $k_\mathrm{min} = 10^{-6} \, \impc$ and $k_\mathrm{max}=1.0 \, \impc$, but allow their amplitudes $A_\mathrm{min}$ and $A_\mathrm{max}$ to vary, with logarithmic prior, in the ranges $-2 <  A < 4$, where $A = \log(10^{10} \Delta^2_\zeta)$.

  \item  Add a ``knot'' with logarithmic prior in $k$, between $k_\mathrm{min} < k_\mathrm{knot} < k_\mathrm{max}$ and allow its amplitude $A_\mathrm{knot}$ to vary in the same ranges as the endpoints in Step 1.

  \item  Interpolate between the endpoints and the knot with linear-spline interpolation.

\end{enumerate}

\begin{figure}
\centering
\includegraphics{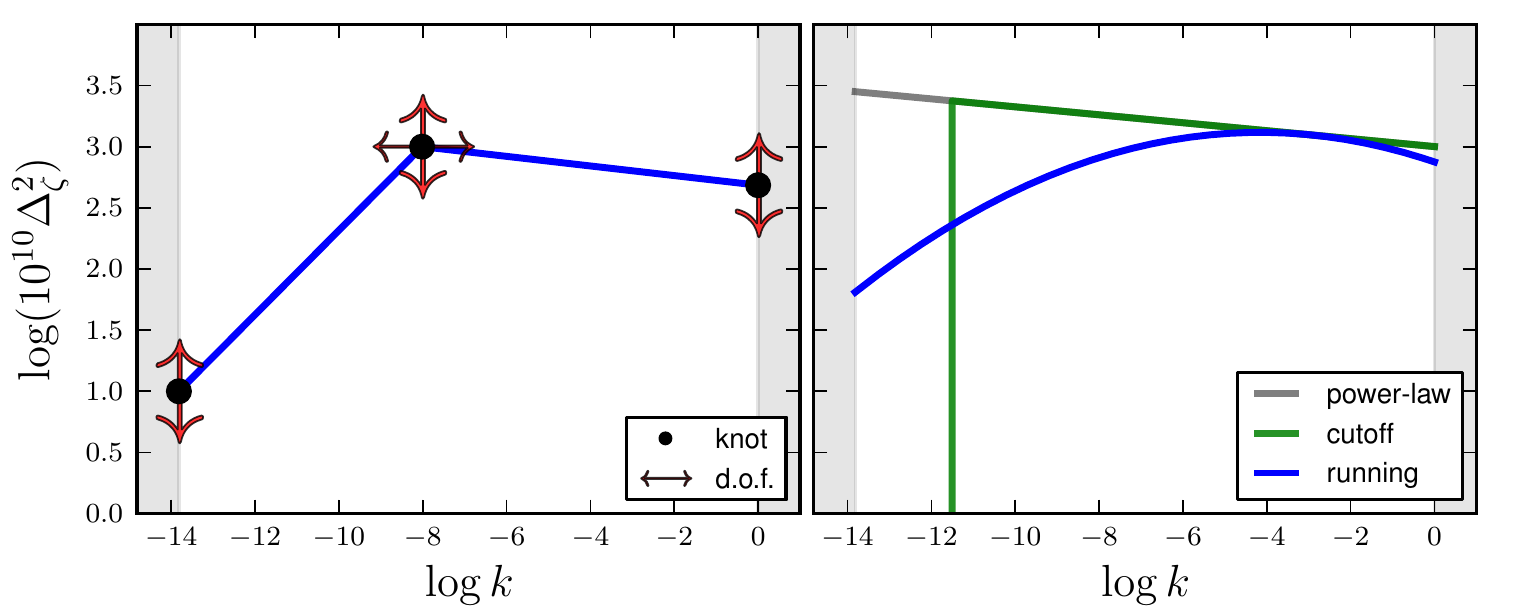}
\caption{\label{model_fig} (\emph{Left}) The knot-spline model for the primordial scalar power spectrum with one knot.  The interior knot adds two degrees of freedom, while the endpoints add one each, for a total of four. (\emph{Right}) The standard power-law, the power-law cutoff at scale $k_\mathrm{cut}$, and a power-law with running.
}
\end{figure}

Varying the knot-location
compensates for the ``look-elsewhere'' effect, as the knot can move over the whole range of $k$.  This permits the reconstruction of features in the scalar primordial power spectrum that could appear at any scale in $k$.  The broad ranges on the priors for $k_\mathrm{knot}$ and $A_\mathrm{knot}$ indicate that possible features are not restricted to large scales.  We call this the ``broad'' prior on the knot's position and amplitude and it gives  conservative values for the Bayesian evidence.

We can also use the information gained during our  \emph{Planck}-only analysis \cite{Aslanyan:2014mqa} to update the prior on the knot position and amplitude.  The \emph{Planck} data indicates that features should only appear at large scales with $k_\mathrm{knot} \lesssim 10^{-3} - 10^{-2} \, \impc$. This reduces the (logarithmic) range of the knot position to two-thirds of the original volume. Furthermore, we know the spectral index is red at large scales, and a low value of $A_\mathrm{knot}$ will generate a blue spectrum  at $k\gtrsim 10^{-3} \, \impc$, so we can further stipulate that $2 < A_\mathrm{knot} < 4$.  This gives the {\em Informative Prior} for the broken spectrum, which yields less conservative and more significant Bayesian evidences. The Planck likelihood is almost zero in the excluded regions, so the  evidence for the informative prior relative to the broad prior is scaled by the ratio of the relative parameter space volume, giving an increase of $\Delta \log Z \approx 1.5$. In what follows we report evidence values with both priors.

We also analyze the standard power-law spectra with both (a) a sharp cutoff at $k_\mathrm{cut}$ and (b) a running spectral index, defined by
\begin{equation}
  \Delta^2 (k) = A_s \left( \frac{k}{k_*} \right)^{n_s -1 + \frac{1}{2} \alpha_s  \log \frac{k}{k_*} } \, ,
  \label{eqn:XXX}
\end{equation}
where $\alpha_s = dn_s/d \log k$ is the running. A running spectral index was considered in the BICEP2 analysis \cite{Collaboration:2014wq}; we repeat this analysis both as a check on our inclusion of the BICEP2 results in our likelihood and in order to calculate Bayesian evidence for the running index.

These two forms of the power spectrum are also illustrated in Fig.~\ref{model_fig}. The pivot scale for the power law  is $k_*=0.05\,\impc$. The cutoff prior is from a logarithmic prior,  $k_\mathrm{cut} \in [1.0\times10^{-6}, 1.0]\,\impc$ and we have a uniform prior on $\alpha_s$ with the range $\alpha_s \in [-0.1, 0.1]$. Note that when the running becomes large  the Taylor expansion in Eq.~\eqref{eqn:XXX} can be an inaccurate parametrization for the inflationary power spectrum~\cite{Abazajian:2005dt}.  As above, we also use an \emph{Informative Prior} for the cutoff spectrum based on the analysis in Ref.~\cite{Aslanyan:2014mqa} with $k_\mathrm{cut} < 10^{-2} \, \impc$.  We report the evidences for both priors, although this makes little difference to the conclusions.

\section{Results}

\begin{table}
\renewcommand{\arraystretch}{1.5}
\setlength{\arraycolsep}{5pt}
\centering
\begin{tabular}{c | c c c}
 \hline
  \hline
  Model & $\Delta\log Z_\mathrm{Broad}$ & $\Delta\log Z_\mathrm{Informative }$ & $2\Delta \log \mathcal{L}_\mathrm{max}$  \\
  \hline
  No Knots & --- & --- & ---\\
  $1$ Knot & $1.6$ & $3.1$ & $6.2$  \\
  \hline
  \hline
  Model & $\Delta\log Z_\mathrm{Broad}$ & $\Delta\log Z_\mathrm{Informative }$& $2\Delta \log \mathcal{L}_\mathrm{max}$ \\
  \hline
  $\Lambda\mathrm{CDM}+r$ & --- & --- & ---\\
                   Cutoff & $0.2$ & $0.6$ & $1.9$  \\
                  Running & $1.1$  & ---& $3.8$ \\
                   \hline
  \hline
\end{tabular}
\caption{Increase in Bayesian evidence, $\Delta\log Z$, and best-fit likelihood, $2 \Delta \log\mathcal{L}_\mathrm{max}$, compared to the standard power law case without running. The reported likelihoods $\mathcal{L}_\mathrm{max}$ are the product of the \emph{Planck} and BICEP2 likelihoods.
}
\label{evidence_table}
\end{table}

\begin{figure}
\centering
\includegraphics[width=1.0\textwidth]{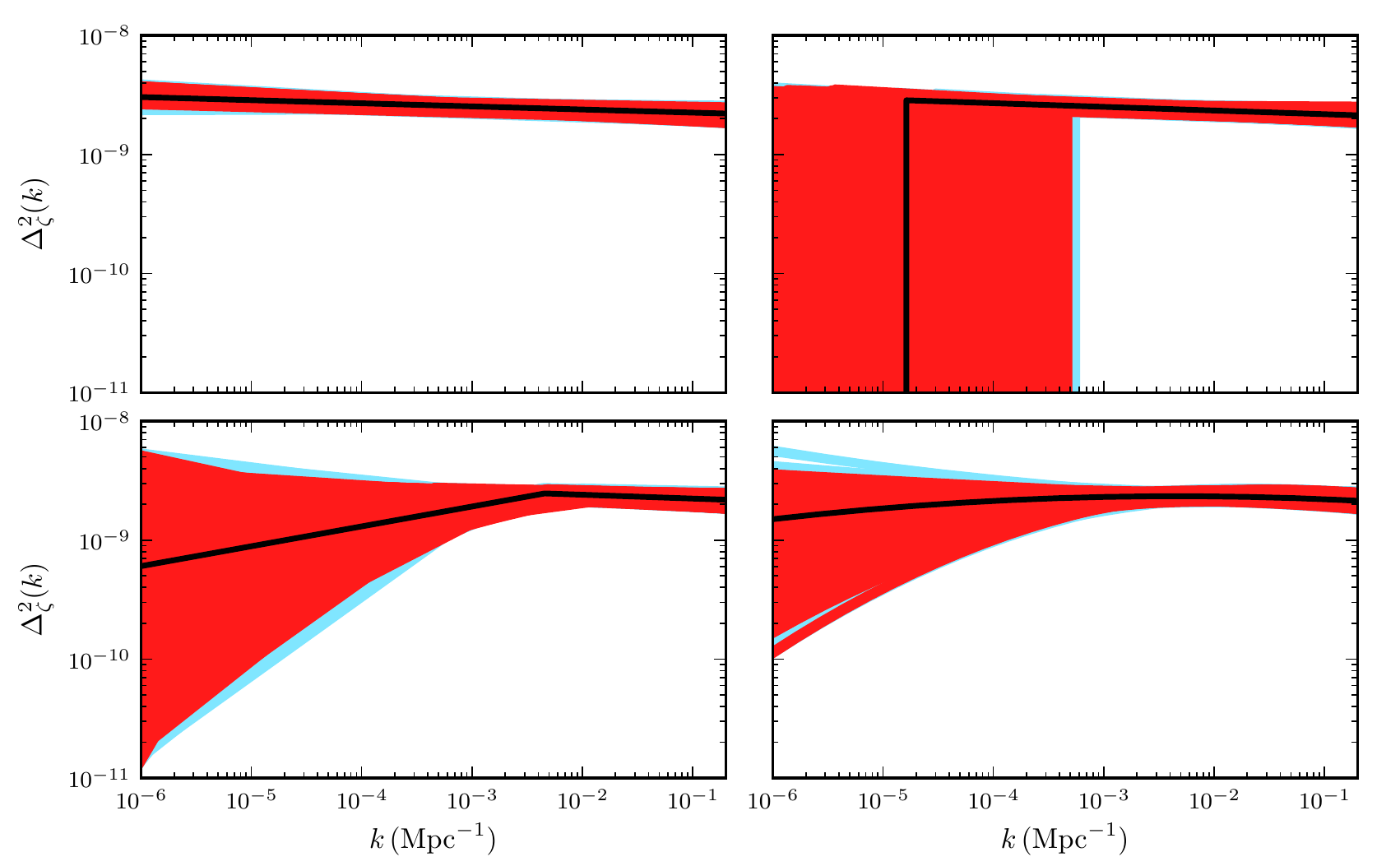}
\caption{\label{pk_lin.fig} (\emph{Left column}) The reconstructed primordial power spectrum with (\emph{top}) $0$ and (\emph{bottom}) $1$ knot. (\emph{Right column}) The power-law spectrum with a (\emph{top}) cutoff and (\emph{bottom}) running. The black solid lines show the best-fit, the red lines are the $68\%$ CI, and the light blue lines are the $95\%$ CI.
}
\end{figure}

\begin{figure}
\centering
\includegraphics[width=0.75\textwidth]{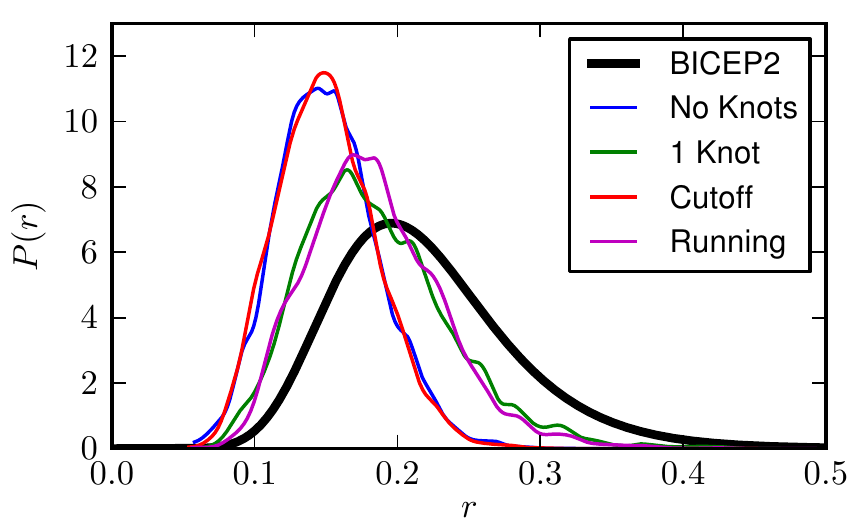}
\caption{\label{r.fig} The posterior distributions for the tensor-to-scalar ratio $r$ from the reconstruction with different models for the scalar power spectrum. For comparison, we also show the BICEP2~\cite{Collaboration:2014wq,Collaboration:2014xx} likelihood for $r$.
}
\end{figure}

\begin{figure}
\centering
\includegraphics[width=0.75\textwidth]{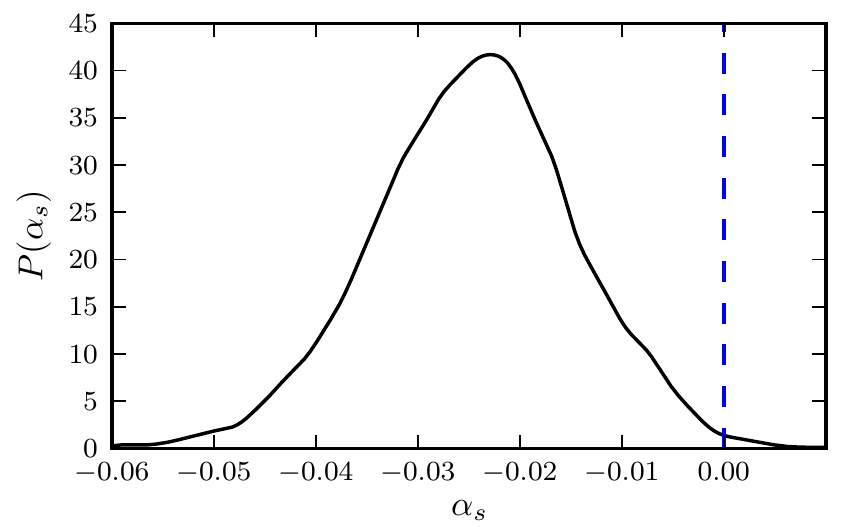}
\caption{\label{running.fig} The posterior distribution for the running of the spectral index.
}
\end{figure}

\begin{table}
\renewcommand{\arraystretch}{1.5}
\setlength{\arraycolsep}{5pt}
\centering
\small
\begin{tabular}{c | c c c c c}
  \hline
  \hline
   & $\Lambda$CDM & No knots & $1$ knot & Cutoff & Running  \\
  \hline
 $\Omega_bh^2$ & $0.02225_{-0.00030}^{+0.00031}$ & $0.02232_{-0.00033}^{+0.00031}$ & $0.02225_{-0.00030}^{+0.00031}$ & $0.02223_{-0.00030}^{+0.00032}$ & $0.02266_{-0.00035}^{+0.00038}$ \\
 $\Omega_ch^2$ & $0.1174_{-0.0029}^{+0.0029}$ & $0.1167_{-0.0029}^{+0.0030}$ & $0.1175_{-0.0030}^{+0.0030}$ & $0.1173_{-0.0028}^{+0.0028}$ & $0.1162_{-0.0032}^{+0.0032}$ \\
 $h$ & $0.690_{-0.014}^{+0.014}$ & $0.693_{-0.014}^{+0.014}$ & $0.689_{-0.014}^{+0.014}$ & $0.690_{-0.013}^{+0.013}$ & $0.698_{-0.016}^{+0.016}$ \\
 $\tau$ & $0.092_{-0.028}^{+0.030}$ & $0.100_{-0.028}^{+0.028}$ & $0.101_{-0.030}^{+0.031}$ & $0.094_{-0.028}^{+0.027}$ & $0.118_{-0.033}^{+0.035}$ \\
 $r$ & $0.150_{-0.032}^{+0.036}$ & $0.149_{-0.033}^{+0.038}$ & $0.178_{-0.044}^{+0.057}$ & $0.149_{-0.034}^{+0.037}$ & $0.180_{-0.043}^{+0.050}$ \\
 $n_s$ & $0.9679_{-0.0084}^{+0.0086}$ & --- & --- & $0.9682_{-0.0082}^{+0.0083}$ & $0.9668_{-0.0096}^{+0.0099}$ \\
 $A_s$ & $3.086_{-0.051}^{+0.054}$ & --- & --- & $3.090_{-0.051}^{+0.048}$ & $3.143_{-0.062}^{+0.063}$ \\
 $\alpha_s$ & --- & --- & --- & --- & $-0.024_{-0.010}^{+0.010}$ \\
 $\log k_\mathrm{cut}$ & --- & --- & --- & $-10_{-2}^{+2}$ & --- \\
 $A_\mathrm{min}$ & --- & $3.428_{-0.064}^{+0.069}$ & $1.89_{-1.50}^{+0.74}$ & --- & --- \\
 $A_\mathrm{max}$ & --- & $3.010_{-0.072}^{+0.071}$ & $3.003_{-0.077}^{+0.078}$ & --- & --- \\
 $\log k_\mathrm{knot}$ & --- & --- & $-5.38_{-5.12}^{+0.69}$ & --- & --- \\
 $A_\mathrm{knot}$ & --- & --- & $3.212_{-0.063}^{+0.107}$ & --- & --- \\
  \hline
  \hline
\end{tabular}
\caption{$68\%$ CI parameter constraints from the reconstruction with different models for the scalar power spectrum. The pivot scale is $k_*=0.05\,\impc$. All of the $A$ values are $A=\log(10^{10} \Delta^2)$.
}
\label{params_table}
\end{table}

Figure~\ref{pk_lin.fig} shows the reconstructed scalar power spectrum with a standard power-law ($0$ knots); a power-law with a sharp cutoff; a power-law with a running spectral index; and a $1$-knot linear-spline model.  The Bayesian evidences are given in Table~\ref{evidence_table}, along with the improvements in best fit improvements for each case. Although the linear spline with no knots is equivalent to the standard power law case without running, the different parameterisations lead to different prior volumes.  Consequently, we report the Bayesian evidence for the cutoff and running power spectra relative to the standard power law case with common priors for $A_s$ and $n_s$, while we report the Bayesian evidence for the 1-knot model compared to the 0-knot power-law model.

In Figure~\ref{r.fig} we show the posterior distributions for $r$ obtained with these  models, along with the result obtained with the BICEP2 data alone.  The posterior distribution for $\alpha_s$ derived for the running power spectrum is shown in Figure~\ref{running.fig}. Constraints for the standard cosmological parameters and the power spectrum parameters are displayed in Table~\ref{params_table}.{\footnote{Although the no knots model is completely equivalent to the standard $\Lambda$CDM model in its functional form, the parametrizations and subsequently the priors are different. For this reason the resulting posteriors on all of the parameters could be slightly different, and we present the results for both cases.}}  As noted previously while the  spline model with no knots is equivalent to $\Lambda$CDM, the different priors associated with the power spectrum parameterisations lead to small differences in ranges for the standard cosmological parameters. Our $68\%$ CI bounds for $\alpha_s$ agree well with those reported in the BICEP2 analysis ($\alpha_s=-0.022\pm0.010$) \cite{Collaboration:2014wq}.

The posteriors for the reconstructed power spectrum in Fig.~\ref{pk_lin.fig} all recover the standard power-law form at small scales $k \gtrsim 10^{-3} \, \impc$.  This confirms the \emph{Planck}-only analysis of Ref.~\cite{Aslanyan:2014mqa}, indicating that the BICEP2 detection of $r\sim0.2$ does not further imply power spectrum features at intermediate to small scales.

At larger scales ($k \lesssim 10^{-3} \, \impc$) all  the non-minimal models indicate a suppression of power in the spectrum of scalar perturbations.  The black lines in Fig.~\ref{pk_lin.fig} are the most likely power spectra, and these all decrease at small $k$.  The increase in the likelihood for the best-fit power spectra are reported in Table~\ref{evidence_table}, although we caution that cosmic variance is important at these scales.  Also, while a local feature at $k\sim 10^{-3} \impc$ could also yield posteriors with large-scale power suppression (as shown in Section~4 of Ref.~\cite{Aslanyan:2014mqa}), we can be more certain about the posteriors in Fig.~\ref{pk_lin.fig}: since the tensor contribution to $C_l^{TT}$ is nearly uniform for scales $l\lesssim 100$, offsetting this increase in power should require a compensating decrease at all scales and not a local feature at intermediate scales. 

The Bayesian evidences in Table~\ref{evidence_table} show some support for the 1-knot linear-spline model and a running spectrum with $\Delta \log Z = 1.6$  and $1.1$,  respectively.  With the \emph{Planck} temperature and lensing data alone, the evidence for the 1-knot model is only $\Delta \log Z=0.7$ \cite{Aslanyan:2014mqa}, indicating that BICEP2 data contributes significantly to the increased evidence.  While these models give qualitatively similar spectra, the running power-law requires that there is less power at scales $k \sim 10^{-3} \, \impc$ in order to achieve the same suppression of large-scale power as the 1-knot model.  The $C_l^{TT}$ at these scales are well-explained by scalar contributions only, but the increased likelihood due to large-scale suppression is partially off-set by the \emph{Planck} likelihood at $l \sim 15$.  This increases the Bayesian evidence for the 1-knot model compared to the running.  The cut-off model gives little overall improvement to the data, since a cut-off causes a large decrease in the $C_l^{TT}$ over all scales larger than the cut-off scale, as seen in Fig.~\ref{cl_tt_fig}.  The 1-knot model generalizes the cut-off at large scales, and thus gives higher evidence values.

Overall, the Bayesian evidences with the broad, uninformative priors show only a mild increase over the power-law $\Lambda\mathrm{CDM}+r$ prediction.  The  evidence computed for the 1-knot model can be characterized as either ``weak'' or ``substantial,'' depending whether one uses the Jeffreys or cosmology scale, as described in Table~\ref{jeffreys_table}.  However, the Informative Prior obtained by incorporating the insight gained from our analysis of the \Planck\ data on its own has a significantly higher evidence due to the reduced parameter volume. This increases $\log Z$ by $\sim 1.5 $, and the resulting evidence of 3.1 constitutes ``strong'' or ``significant'' evidence for the suppression of power in the spectrum of primordial perturbations at large scales.

\section{Discussion}

Using data from  \emph{Planck} and BICEP2 we perform parameter estimates and calculate  Bayesian evidence in order to explore the implications of the BICEP2 result for different parameterisations of the primordial scalar power spectrum. In agreement  with the BICEP2 analysis we find that the posterior probability for a running spectral index excludes zero at the 95\% confidence interval. Similarly, a spectrum defined by a linear spline with one internal knot shows a distinct preference for a suppression of power in the scalar spectrum at large angular scales, $k \lesssim 10^{-3} \, \impc$. For an informative prior incorporating the results of the \Planck-only analysis \cite{Aslanyan:2014mqa} the corresponding Bayesian model selection criteria show a pronounced preference for a model with broken scale invariance, with $\Delta \ln Z = 3.6$.

This paper extends the analysis of  Ref.~\cite{Aslanyan:2014mqa}, which reconstructs the primordial scalar power spectrum from \emph{Planck} temperature data alone.  In particular, Ref.~\cite{Aslanyan:2014mqa}  shows that the evidence for  extra structure in the scalar power spectrum is negligible and that this reconstruction technique successfully recovers artificial signals injected into simulated temperature maps.  Consequently, these  results quantify the impact of the additional information provided by BICEP2.

If the scalar power spectrum is not well described by the usual power-law form, the estimated values of other novel cosmological parameters may also be modified, including sum of the neutrino masses or the number of effective relativistic degrees of freedom, which can be degenerate with features in the scalar power spectrum \cite{2013PhRvD..87l3528H,2011NJPh...13j3024S,2003MNRAS.342L..72B,2002PhRvD..66j3508T,2001PhRvD..63d3001K,dePutter:2014vd}.  Likewise, the BICEP2  data does not put robust constraints on the tensor index $n_T$.  Generically, the expectation from inflation is that $n_T=-r/8$ to first order in slow-roll, so we assume a scale-invariant tensor spectrum in this analysis. However, while a blue (or sharply peaked) tensor background would also  alleviate the tension between \emph{Planck} and BICEP2, the physical processes that generated this spectrum would at least be as radical as considering a non--power-law scalar spectrum.

The recent BICEP2 result provides strong evidence for a primordial tensor background, from which it is inferred that the very early universe underwent an inflationary phase. However, the results presented here  imply that BICEP2 also suggests  that this inflationary phase yields a non-trivial scalar power spectrum, and that the underlying inflationary mechanism is not well-described by a simple, smooth single-field potential. The inverse problem associated with reconstructing the inflationary potential from data has been widely discussed \cite{Copeland:1993ie, Copeland:1993jj,Lidsey:1995np,Hansen:2001eu,Leach:2002ar, Leach:2002dw, Leach:2003us,Kinney:2006qm,Peiris:2006ug,Peiris:2006sj,Martin:2006rs,Lesgourgues:2007gp, Lesgourgues:2007aa, Peiris:2008be,Adshead:2008vn,Hamann:2008pb,Kinney:2008wy,Martin:2010hh,Easther:2011yq} and these methods would have at least three nontrivial input parameters in such a scenario.   Likewise, with a large negative running similar to the central value found here, simple inflationary models typically yield an unacceptably small number of e-foldings, implying that the running itself must be scale dependent \cite{Easther:2006tv}.

Needless to say, this analysis takes both the current \Planck\ and BICEP2 data-products at face-value.   The most conservative explanation for these result is that future analyses will eliminate much of the apparent tension between BICEP2 and other cosmological datasets. From this perspective our analysis quantifies the extent of that tension.

\acknowledgments

We acknowledge the contribution of the NeSI high-performance computing facilities and the staff at the Centre for eResearch at the University of Auckland, especially Mark Gahegan and Gene Soudlenkov. New Zealand's national facilities are provided by the New Zealand eScience Infrastructure (NeSI) and funded jointly by NeSI's collaborator institutions and through the Ministry of Business, Innovation \& Employment's Research Infrastructure programme [\url{http://www.nesi.org.nz}]. KNA is supported by NSF CAREER Grant No. PHY-11-59224.

\bibliographystyle{JHEP}
\bibliography{references}

\end{document}